\documentclass[aps,pre,twocolumn,%linenumbers
]{revtex4-1}
\usepackage{epsfig}
\usepackage{hyperref}
\usepackage{amsfonts}
\usepackage{amsmath}
\usepackage{color}
\usepackage{mathtools}
 
\newcommand{\HL}[1]{\textcolor{red}{\sf #1}}

\usepackage{graphicx}
\newcommand\sbullet[1][.5]{\mathbin{\vcenter{\hbox{\scalebox{#1}{$\bullet$}}}}}

\begin{document}

\title{Active noise-driven particles under space-dependent friction in one dimension}

\author{D. Breoni$^1$, H. L\"owen$^1$, and R. Blossey$^2$} 
\address{$^1$Institut f\"ur Theoretische Physik II: Weiche Materie,
Heinrich Heine-Universit\"at D\"usseldorf, Universit\"asstra\ss e 1, 40225 D\"usseldorf, Germany\\
$^2$University of Lille, UGSF CNRS UMR8576, 59000 Lille, France}  

\begin{abstract} We study a Langevin equation describing the stochastic motion of a particle in one dimension with coordinate $x$,
which is simultaneously exposed to a space-dependent friction coefficient $\gamma(x)$, a confining potential $U(x)$ and 
non-equilibrium (i.e., active) noise. Specifically, we consider frictions $\gamma(x)=\gamma_0 + \gamma_1 | x |^p$ 
and potentials $U(x) \propto | x |^n$ with exponents $p = 1,2$ and $n = 0, 1, 2$. We provide analytical and numerical 
results for the particle dynamics for short times and the stationary PDFs for long times.
The short-time behavior displays diffusive and ballistic regimes while the stationary PDFs
display unique characteristic features depending on the exponent values $(p, n)$. The PDFs interpolate between 
Laplacian, Gaussian and bimodal distributions, whereby a change between these different behaviors can be achieved by a tuning 
of the friction strengths ratio $\gamma_0/\gamma_1$. Our model is relevant for molecular motors moving on a one-dimensional 
track and can also be realized for confined self-propelled colloidal particles. 
\end{abstract}

\maketitle

\section{Introduction}

Particles moving under the influence of a stochastic driving force in one dimension \cite{bouchaud_classical_1990} are a fruitful laboratory for the exploration of the 
statistical mechanics of active systems, since they allow, in suitably chosen cases, for an analytic treatment. Following the initial works
on one-dimensional active particles  \cite{tailleur_statistical_2008,lindner_diffusion_2008}, the problem is currently receiving increased attention, since the results can 
be of relevance for various soft matter and biological systems in a larger sense \cite{toner_hydrodynamics_2005,ramaswamy_mechanics_2010,cates_motility-induced_2015,bechinger_active_2016,elgeti_physics_2015}. 
One-dimensional models for active particles, 
in spite of their inherent simplicity, are indeed of relevance even for the description of collective effects \cite{romanczuk_collective_2010,ben_dor_ramifications_2019,illien_speed-dispersion-induced_2020,teixeira_single_2021}.

A standard type of model under scrutiny is the persistent Brownian motion, the persistence being forced by activity.
Maybe the simplest model for an active particle in one dimension is a discrete {\it run-and-tumble process}  where the direction of self-propulsion 
 discretely flips, i.e.\   the driving is assured by a random directional velocity, see, e.g. 
\cite{demaerel_active_2018,malakar_steady_2018,dhar_run-and-tumble_2019,ben_dor_ramifications_2019,le_doussal_velocity_2020,bialas_colossal_2020,dean_2021,mori_universal_2020}. 

It is defined by the Langevin equation
\begin{equation}
\dot{x}(t) = v_0 \sigma(t) 
\end{equation}
where the stochastic term  $\eta(t) = v_0 \sigma(t) $ is a telegraphic noise with values $\pm v_0$, with the sign flipped at a given tumbling rate.
In particular, this model has been explored for a single particle in the presence of external potentials 
\cite{angelani_confined_2017,razin_generalized_2017,razin_entropy_2020} and random disorder  \cite{ben_dor_ramifications_2019,le_doussal_velocity_2020}.

On a second level of complexity, one can consider a Brownian particle self-propelled along its orientation such that only the projection on the $x$-axis is contributing to the actual particle propulsion but the orientation diffuses on the unit circle or unit spheres \cite{ten_hagen_brownian_2011}. 
These models of {\it active Brownian particles}
were extensively discussed in the literature \cite{bechinger_active_2016} and can be realized by self-propelled Janus-colloids 
in channel-like confinement \cite{wei_single-file_2000,lutz_single-file_2004,herrera-velarde_ordering_2010}.
For low activity, the fluctuation-dissipation theorem which couples the strength of the Brownian noise and the friction via the bath
temperature should be fulfilled. Hence, in the limit of vanishing activity, the stationary probability 
density function (PDF) is a Boltzmann distribution. Also simpler variants of these models where the drive
 just enters via colored noise, often called active Ornstein-Uhlenbeck particles have been explored in one dimension 
 \cite{szamel_self-propelled_2014,Wittmann_2017,Das_2018,caprini_2018,Caprini_2018-2,sevilla_generalized_2019}.

A third complementary approach starts from Langevin equations coupling an active white noise term to 
a spatially dependent diffusion coefficient \cite{cherstvy_anomalous_2013}, or friction \cite{kumar_active_2008,baule_exact_2008}. The basic idea here is
 the gradient in the friction induces a drift velocity which drives the particle at constant noise.
In near-equilibrium situations, a spatial dependence of the friction enforces a spatial dependence 
of the noise strength according to the fluctuation-dissipation 
theorem which guarantees a relaxation of the PDF to the stationary Boltzmann distribution. 
Here we deliberately abandon the validity of the fluctuation-dissipation 
theorem and therefore postulate a {\it non-equilibrium noise} in the presence of a friction gradient
to define a nonequilibrium model with inherent activity. We refer to this kind of noise as ``active" noise in the sequel.
The equilibrium limit of a stationary Boltzmann 
distribution is reached if the friction gradient vanishes. 
Though these kind of  non-equilibrium noise models 
were proposed more than a decade ago
\cite{kumar_active_2008,baule_exact_2008} and bear interesting descriptions for the
 biologically motivated case 
of molecular motors moving on a one-dimensional track \cite{mogilner_motion_1998,fogedby_exact_2004,kolomeisky_dynamic_2005,makhnovskii_reciprocating_2006,rozenbaum_reciprocating_2010,makhnovskii_fluctuation-induced_2014}
such as the action of chromatin remodeling motors on nucleosomes \cite{blossey_chromatin_2019}, they have not yet 
been studied systematically.

Here we propose  a class of one-dimensional models with active noise in different 
friction gradients and external confining potentials which we solve analytically. Our motivation 
to do so is threefold: first, any exactly soluble model in nonequilibrium is of fundamental importance for a basic understanding
of particle transport. Second, we obtain qualitatively different PDFs which can be categorized within these active noise models. 
Third, our results are relevant for applications in the biological context and for artificial colloidal particles.

The model we discuss is based on a Langevin equation of a particle with nonequilibrium noise and space-dependent friction
in one dimension with a spatial coordinate $x$. The particle is exposed to
a space-dependent friction coefficient $\gamma(x)=\gamma_0 + \gamma_1 | x |^p$ and  an external potential $U(x)\propto | x |^n$
with exponents $p = 1,2$ and $n = 0, 1, 2$. For short times, we provide analytical results for the MD and the 
MSD. Depending on the parameters, we find
 a crossover from an initial diffusive to a ballistic regime  for $p=1,2$ and $n\neq 0$ 
as typical for any 
model of a single free active particle. For long times and $n>0$, we obtain the stationary 
probability density functions (PDFs) from 
the corresponding Fokker-Planck equation. The PDFs are non-Boltzmannian and 
display a rich variety of behaviors: from Gaussian-like to Laplace-like distributions,
and variants of bimodal-Gaussian like distributions. A  change between 
these different behaviors can be achieved by a tuning of the ratio of the friction parameters $\gamma_0/\gamma_1$.  
To test the robustness of our results, we evaluate the effect of additional thermal noise 
\cite{kumar_active_2008,baule_exact_2008}.  

As already mentioned, our proposed model is relevant for molecular motors moving on a one-dimensional track
and can also be realized for confined self-propelled colloidal particles. In fact, colloids can be exposed to almost any 
arbitrary external potential by using optical fields \cite{evers_colloids_2013,lozano_phototaxis_2016,jahanshahi_realization_2020} 
and almost any  kind of noise can externally be programed by external fields
\cite{fernandez-rodriguez_feedback-controlled_2020,sprenger_active_2020}. A space-dependent friction can be imposed be a viscosity gradient in the suspending medium on the particle scale, 
a situation typically encountered for viscotaxis \cite{liebchen_viscotaxis_2018,stehnach_viscophobic_2020,shirke_viscotaxis-_2019,lopez_dynamics_2020}.
 
\section{A particle under nonequilibrium noise: The model} 

Following \cite{baule_exact_2008}, the model Langevin equation of a single active particle  
on a one-dimensional trajectory $x(t)$  we use in this work is given by the expression
\begin{equation} \label{active}
\gamma(x) \dot{x}(t) = - U'(x) + \sqrt{A} \xi(t)\, 
\end{equation}
in which $U(x)$ is the confining potential, and $\xi(t)$ a Gaussian random noise with
\begin{equation}
\langle \xi(t) \rangle = 0\,,\,\,\, 
\langle \xi(t)\xi(t') \rangle = \delta(t-t')\, 
\end{equation}
and $A>0$ characterizes the noise strength. The brackets $<...>$ denote a noise-average.
The Langevin equation (\ref{active}) can be rewritten in the standard multiplicative noise form as
\begin{equation} \label{active2}
\dot{x}(t) = - \frac{U'(x)}{\gamma(x)} + \frac{\sqrt{A}}{\gamma(x)} \xi(t)\, ,
\end{equation}
which we will interpret in the Stratonovich sense. 

\begin{figure}
	\includegraphics[scale=0.38]{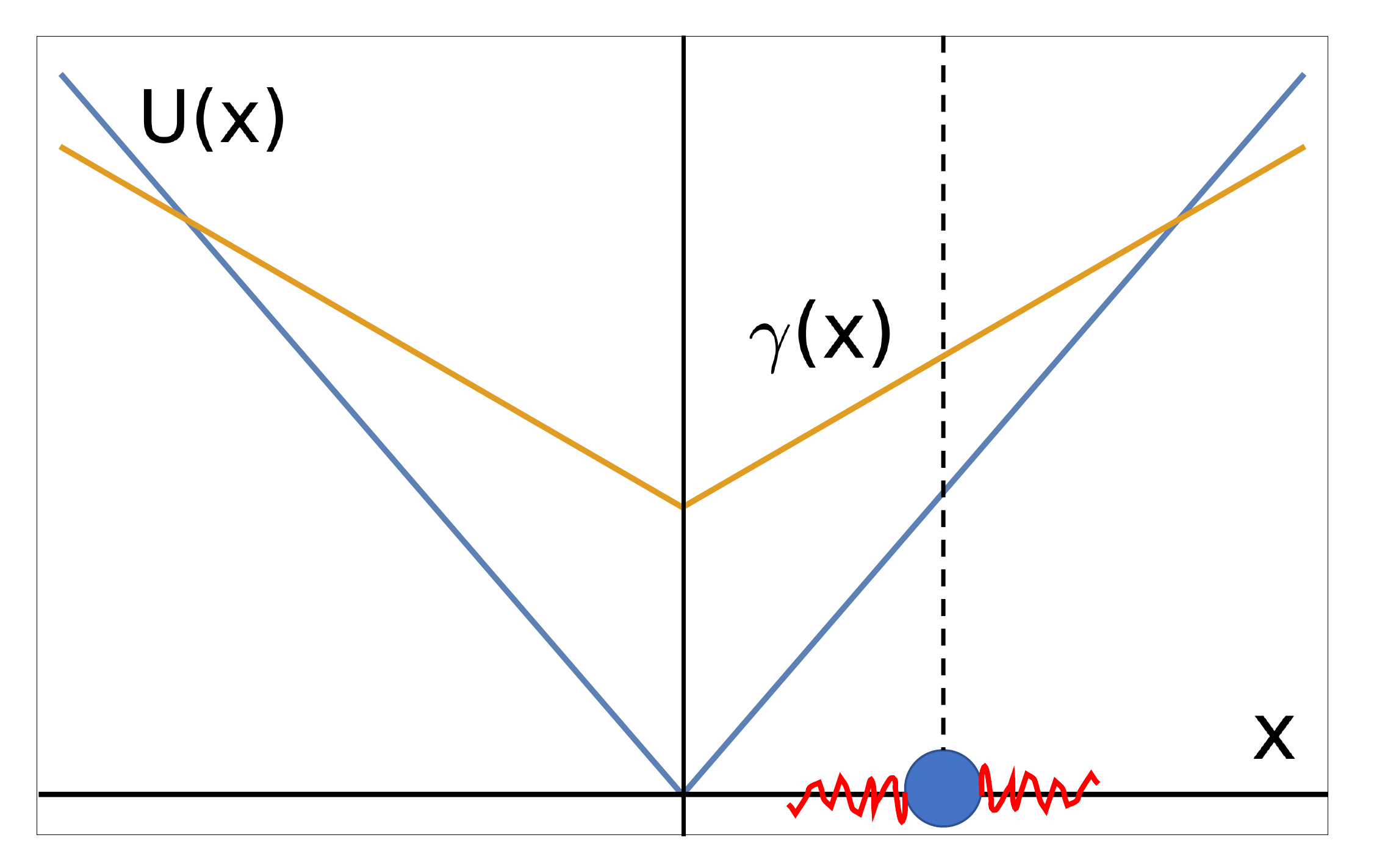}
	\caption{Sketch of the confining potential $U(x) = \kappa |x|$, a linear friction gradient 
 	$\gamma(x) = \gamma_0 + \gamma_1 |x|$ in arbitrary units.The particle, shown by a blue dot on the
	$x$-axis, is activated by noise (indicated in red), under the influence of the potential and the friction gradient.}
         \label{F1}
\end{figure}

The factor $\gamma(x) $ in Eqs.(\ref{active}), (\ref{active2}) is a space--dependent friction force. It has been introduced in models for
molecular motors in \cite{kumar_active_2008} and been modeled by an expression $ \gamma(x) = 1 + \delta \tanh(x\beta) $ 
with parameters $\delta$, $\beta$ ($0< \delta <1$), a function saturating at both large positive 
and negative values of the argument displaying a linear crossover zone. Aiming at analytic results, in this work we use an algebraic expression   
\begin{equation}
\gamma(x) = \gamma_0 + \gamma_1 |x|^p\, 
\end{equation}
for the friction term with two parameters $\gamma_0>0$ and $\gamma_1\geq 0$ and an integer exponent $p\geq 0$, which, although unbounded, will allow us to uncover interesting properties of the stationary probability density functions.   
These arise when we consider the particle in low-order polynomial confining potentials which we take to be of the general form
\begin{equation}
U(x) = \frac{\kappa}{n}|x|^n\, 
\end{equation}
with $\kappa \geq 0$ and another integer exponent $n\geq 0$
An illustration of the situation we address is given for the case $p=n=1$ corresponding to a wedge-like potential $U(x) = \kappa |x|$ 
with a friction term $\gamma(x) = \gamma_0 + \gamma_1 |x|$, see Fig. \ref{F1}. 

\section{Short-time behavior}

We start our discussion by determining the short-time behavior of the active-noise driven
 particle and compute the short-time mean displacement (MD) and 
the mean-square displacement (MSD) for the Langevin equation (\ref{active}), as done previously \cite{breoni_2020}. 
Specifically, we address the cases of 
a freely moving particle, $U'(x)=0$ (i.e.\ $n=0$) and a particle moving in the potential $ U(x) = (\kappa/n) |x|^n $ for 
$n=1, 2$, which respectively correspond to a particle on a (double) ramp (or, under gravity) and in a harmonic oscillator potential. 

\subsection{Constant friction gradient}

{\bf Free particle.} First we consider the case of $p=1$, i.e. a constant friction gradient acting on a free particle. Due to the spatial dependence
of the friction term, the choice of initial position $x_0=x(t=0)$ is important. In the immediate vicinity of the origin, the initial motion will be that of a free Brownian 
particle since $\gamma_0 \gg \gamma_1 |x_0| $. In order to see an effect of the $x$-dependence of the friction term, we place the particle initially far 
away from the origin with $|x_0| \gg 0 $ to prevent the particle to traverse 
from the positive sector $x_0 > 0$ to the negative sector $ x_0 < 0$ or vice versa, so that we ignore the nonanalyticity of $\gamma(x)$ at the origin.
We can then consider the case $x_0>0$, drop the modulus and use separation of the variables 
in Eq.(\ref{active}) to find
\begin{equation}
\gamma_0(x(t) - x_0) + \frac{\gamma_1}{2}(x(t)^2 - x_0^2) = \sqrt{A}\int_0^t dt '\xi(t') \,
\end{equation}
resulting in
\begin{equation}
x(t) = \gamma_1^{-1}\left(- \gamma_0 + \sqrt{\gamma^2_0 + c(t)}\right)  
\label{quadratic}
\end{equation}
with
\begin{equation}
c(t) \equiv 2\gamma_1\left(\frac{\gamma_1}{2}x_0^2 + \gamma_0 x_0 + \sqrt{A}\int_0^t dt' \xi(t') \right)\, .
\end{equation}
The resulting MD $\langle x(t) - x_0 \rangle$ can then be obtained by an expansion of the square root as
\begin{eqnarray}
\label{Eq10}
\langle x(t) - x_0 \rangle & = & \frac{1}{\gamma_0+\gamma_1x_0} \overline{\xi}(t)\,\,\,  \\
&& \hspace*{-2cm} + \sum_{m=2}^\infty (-1)^{m\HL{-1}} \frac{(2m-3)!}{2^{m-2}m!(m-2)!} \frac{\gamma_1^{m-1}\overline{\xi}^m(t)}{(\gamma_0+\gamma_1x_0)^{2m-1}} \nonumber
\end{eqnarray} 
where
\begin{eqnarray}
&& \hspace*{-10mm} \overline{\xi^m}(t) \equiv \left\langle \left(\sqrt{A} \int_0^t dt' \xi(t')\right)^m \right\rangle  \\
&& \hspace*{20mm} =\begin{cases}
\frac{m!}{2^{m/2}(m/2)!}(At)^{m/2} & m\text{ even}\\
0 & m\text{ odd}\, 
\end{cases} \nonumber
\end{eqnarray}
such that the final expression for the MD, after reintroducing the left side of the plane by symmetry, is
\begin{equation} \label{msd1} 
\hspace{-5cm}\langle x(t)-x_0 \rangle   
\end{equation}   
\vspace{-0.7cm} 
\[
= -\text{sgn}(x_0)\sum_{m=1}^\infty\frac{(4m-3)!}{2^{3m-2}m!(2m-2)!}\frac{\gamma_1^{2m-1}}{\gamma(x_0)^{4m-1}}(At)^m\, .
\] 
The details of how we obtained Eq.(\ref{Eq10}) can be found in the appendix.

Let us now discuss this result for the MD in more detail: first of all, if the friction gradient vanishes 
(i.e., in the case $\gamma_1=0$), there is no drift at all as ensured by left-right symmetry. Second, 
for positive friction gradients
 $\gamma_1$ the leading term for short times in the MD is linear in time 
and in the friction gradient $-\text{sgn}(x_0) \gamma_1 A t / 2 \gamma(x_0)+\mathcal{O}(t^2)$
resulting in a drift velocity of $-\text{sgn}(x_0) \gamma_1 A  / 2 \gamma(x_0)$.
 Interestingly the particle drift is along the negative gradient of the friction implying that the particle migrates on average to the place where the friction is small. This is plausible since at positions with smaller friction there are stronger fluctuations which promote the particle to the position of even lower friction on average. 
A similar qualitative argument was put forward for colloids moving under hydrodynamic 
interactions (see ref. \cite{doi_edwards_1986}, p.54), which represent another case of multiplicative noise, see also \cite{lau_lubensky_2007}.
Third, in a more mathematical sense, 
the series in Eq.(\ref{msd1}) is an asymptotic series which strictly speaking does not converge for $m\to \infty$ 
 but nevertheless gives a good approximation to the MD to any finite order in time. This asymptotic expansion even holds 
if the cusp in the friction at $x=0$ were to be included as any corrections do not contribute 
to the short-time expansion in powers of time.

Similarly, one can calculate the {\it mean-squared displacement} (MSD), which we define as 
\begin{equation} \label{msd1} 
\Delta(t)=\langle (x(t)-x_0)^2 \rangle\, .
\end{equation} 
One obtains a simple relation to the MD as follows
\begin{equation}
 \Delta(t)= -\text{sgn}(x_0)\frac{2\gamma(x_0)}{\gamma_1}\langle x(t)-x_0 \rangle.
\end{equation}

Taking the asymptotic series as an approximation for finite times, we can now discuss 
for both the MD and the MSD the {\it crossing times\/} $t_{m\rightarrow m+1}$, 
defined as the ratios $A_m/A_{m+1}$ between two  consecutive regimes scaling with $A_mt^m$ and $A_{m+1}t^{m+1}$. These crossing times define the moments at which the $(m+1)^{th}$ terms of the time series start to dominate over the previous ones \cite{breoni_2020}. In this case, the crossing times of both the MD and MSD are given  by
\begin{equation}
t_{m\rightarrow m+1}=\frac{4(m+1)(2m-1)}{(4m+1)(4m-1)(4m-2)}\frac{\gamma(x_0)^{4}}{A\gamma_1^{2}}\, .
\end{equation}
The sequence of crossing times is monotonously decreasing, i.e. crossing times between larger regimes always occur before those of smaller ones. 
This in turn means that the only real regime for the free particle is the first one, linear in time. The same reasoning applies to the MSD, as it is proportional to the MD.

Generally, we characterize these regimes with time-dependent {\it scaling exponents}
\begin{equation}
\beta(t) \equiv \frac{d(\log_{10}(\langle x(t)-x_0 \rangle))}{d(\log_{10}(t))}
\end{equation}
and
\begin{equation}
\alpha(t) \equiv \frac{d(\log_{10}(\Delta(t)))}{d(\log_{10}(t))}\, .
\end{equation}
If these exponents are constant over a certain regime of time they indicate that the MD (or the MSD) are a power-law in time
proportional to $t^{\beta}$ (or $t^{\alpha}$).

Finally, we define a typical {\it passage time\/} 
for the particle to reach the origin and cross the cusp in the friction at $x=0$. Beyond such a passage time 
our theory should not be applicable any longer, as we ignored the presence of the cusp in the friction. We decided to run the simulations for longer than this time in order to show how the theory breaks down.
Such a typical passage time $t_1^c$ is set by requiring
\begin{equation} 
\langle x(t_1^c)\rangle \equiv 0\, ,
\end{equation} 
which means that on average the particle has reached the origin. Of course this is only an estimate. The definition
of a passage time can be improved 
by requiring that the particle is one standard deviation away from the origin on average 
\begin{equation}
 \langle x(t_2^c)\rangle+\sqrt{\Delta(t_2^c)} \equiv 0
\end{equation} 
for $x_0>0$. This defines a second typical passage time $t_2^c$ which is in general smaller than $t_1^c$.
Taken together, the two passage times  $t_1^c$ and $t_2^c$ provide a rough estimate for the validity of our theory.

\begin{figure}
	\includegraphics[scale=0.21]{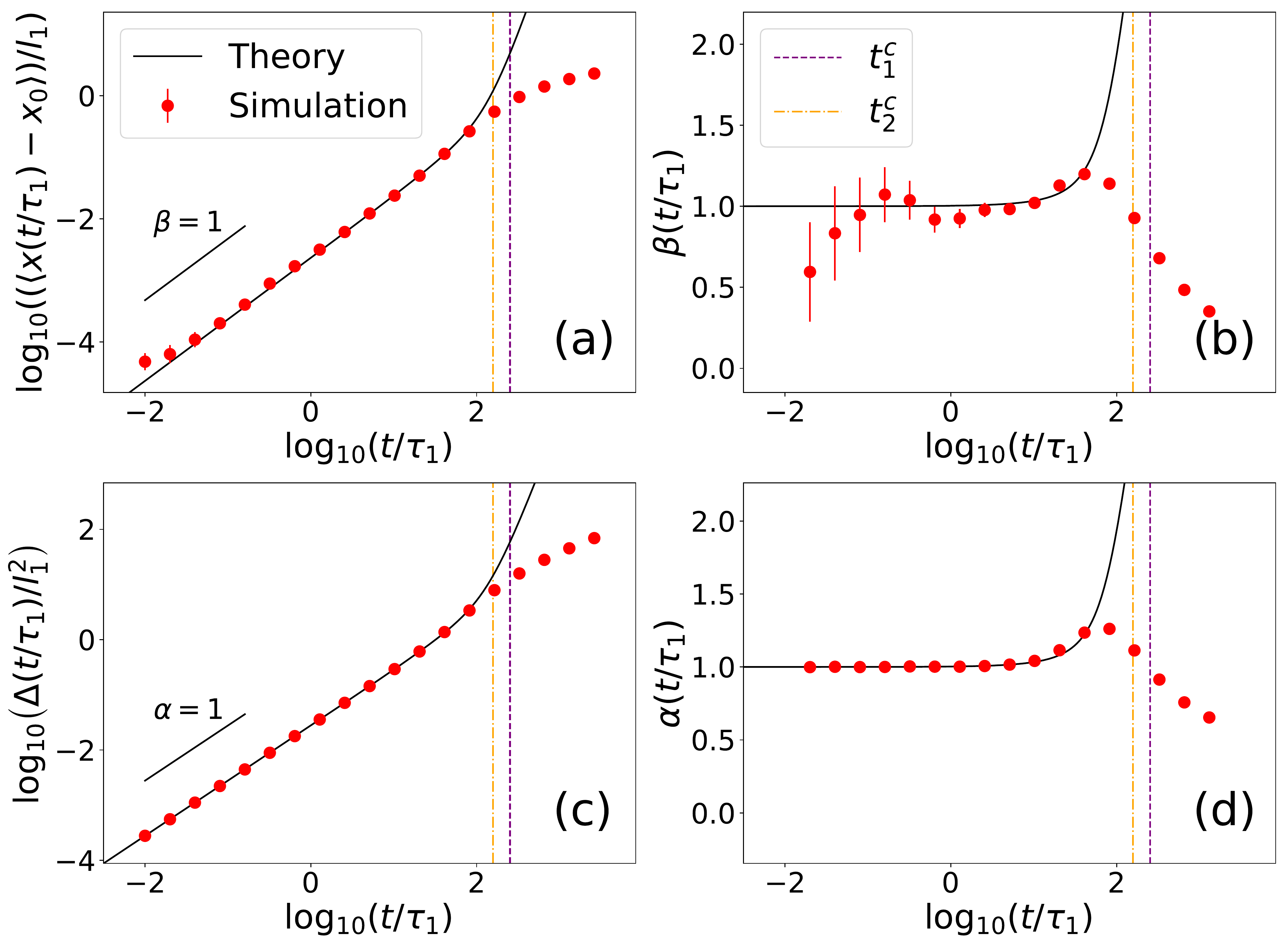}
\caption{Constant friction gradient and free particle $(p,n)=(1,0)$. (a) mean displacement; (b) associated scaling exponent $\beta(t)$; 
(c) mean-squared displacement $\Delta(t)$; (d) associated scaling exponent $\alpha(t)$. 
The length unit is $l_1\equiv \gamma_0/\gamma_1$, while the time unit  is $\tau_1\equiv l_1^2/A$. 
The initial position is $x_0 = 5l_1$. Simulation data are shown with error bars as red symbols. The theory is the solid line. 
The typical passage times $t_1^c$  and $t_2^c$ are indicated by purple and orange vertical lines. 
}
\label{F2}
\end{figure}

Explicit data for the MD and MSD are shown in Fig. \ref{F2} a) and c), with the associated 
exponents $\beta (t)$ and $\alpha (t)$ given in Fig.  \ref{F2} b) and d).  
The typical passage times $t_1^c$ (in purple) and $t_2^c$ (in orange) are also indicated by vertical lines.
In the figure we compare our analytic results (taken by summing up the series up to a finite  order of 5)
with the full numerical solution of the Langevin equation, Eq.(\ref{active2}), in Stratonovich interpretation; 
details of the numerical method are discussed in the Appendix. 

First of all in the time regime $t<t_2^c$ the asymptotic theory is in good agreement with the simulation data.
Both theory and simulations are dominated by the linear time-dependence in the MD and MSD as indicated by the
slope of the MD and MSD and likewise by the scaling exponents $\beta (t)$ and $\alpha (t)$ which are both close to unity.
In both theory and simulation the scaling exponents $\beta (t)$ and $\alpha (t)$ first show a trend to increase to transient values larger 
than unity, i.e. towards superdiffusive behavior. Beyond $t_2^c$ this trend weakens in the simulations 
such that both exponents fall significantly below unity. This is due to the fact that the particle has arrived at the position of minimal friction at the origin and therefore decelerates.
However, in the theory there is an artificial  monotonic increase in the slope due to the fact that there is even unphysical negative frictions for position smaller than $\gamma_0/\gamma_1$ (for the case $x_0>0$).

{\bf Linear confining potential.} Now we consider the case $n=1$ where $U(x)=\kappa |x|$, for $p=1$. 
As before, we assume $x_0 \gg 0$ and drop the modulus in 
the potential. The force is then constant $U'(x)= - \kappa$  and the equation of motion can be solved by
separation of variables as in the free case $n=0$. The result for the MD is
\begin{eqnarray}
\hspace{-.4 cm}\langle x(t)-x_0\rangle & = & -\text{sgn}(x_0)\left[	\frac{\kappa t}{\gamma(x_0)}  +\sum_{m=2}^{\infty}\frac{(2m-3)!}{2^{m-2}(m-2)!} \right. \nonumber \\
&&\hspace*{-1cm} \left.\times\frac{\gamma_1^{m-1}}{\gamma(x_0)^{2m-1}} \sum_{k=0}^{\lfloor m/2 \rfloor}\frac{A^k\kappa^{m-2k}}{(m-2k)!2^kk!}t^{m-k}\right],
\end{eqnarray}
where the Gauss bracket $\lfloor \cdot \rfloor$ indicates the closest integer from below and the case $x_0 < 0$ is reintroduced via left-right symmetry. 
For short times, the MD is given by
\begin{align}
\langle x(t)-x_0\rangle  = &-\text{sgn}(x_0)\left[ \left(\frac{\kappa}{\gamma(x_0)} + \frac{\gamma_1 A}{2 \gamma(x_0)^3}\right)t\right.  \\
&\hspace*{-1cm} \left.+\left(\frac{\gamma_1 \kappa^2}{2\gamma(x_0)^3}+\frac{3\gamma_1^2 \kappa A}{2\gamma(x_0)^5}+\frac{15\gamma_1^2 A^2}{8\gamma(x_0)^7}\right)t^2\right]+\mathcal{O}(t^3) \nonumber
\end{align}
with an initial effective drift velocity
\begin{equation}
-\text{sgn}(x_0) \left(\frac{\kappa}{\gamma(x_0)} + \frac{\gamma_1 A}{2 \gamma(x_0)^3}\right)\, ,
\end{equation}
%comment HL: Davide, can you provide the drift velocity, i.e. the terms linear in t in the MD
%comment HL: Davide, I do not think that the old formula (21) for the MD  is correct on dimensional grounds a length x cannot be kappa times t  a prefactor involving the frictions is missing here
which is a superposition of two effects arising from: 
i)  the direct force $-\text{sgn}(x_0)\kappa$ already present in the equilibrium noise case
(where $\gamma_1=0$), and ii) the linear friction gradient. 
%comment HL: Davide, check whether this is true
As in the free particle case ($n=0$), the MD and the MSD fulfill a linear relationship given by
\begin{equation}
\Delta(t)  = -\frac{2\gamma(x_0)}{\gamma_1}\left[\frac{\kappa t}{\gamma(x_0)} +\text{sgn}(x_0)\langle x(t)-x_0 \rangle\right]\, ,
\end{equation}
such that the short-time expansion for the MSD is given by
%comment HL: Davide, can you give the short-time expansion up to t^2
\begin{eqnarray}
\Delta(t) & =  & \frac{A}{ \gamma(x_0)^2}t+\left(\frac{\kappa^2}{\gamma(x_0)^2}+\frac{3\gamma_1 \kappa A}{\gamma(x_0)^4}+\frac{15\gamma_1 A^2}{4\gamma(x_0)^6}\right)t^2 
\nonumber \\
&& \hspace*{2cm} + \mathcal{O}(t^3). 
\end{eqnarray}
Clearly, for $\kappa = 0$, the free case is recovered. 

We see from the MSD that we have first a diffusive and later a ballistic regime
 while for the MD the dominating term is the drift, 
as the particle feels the effects of the constant force. 
In fact, the crossing time between these two regimes in the MSD is
\begin{equation}\label{t_crossover} 
t_{1\rightarrow 2}=\frac{4A\gamma(x_0)^4}{4\kappa^2\gamma(x_0)^4+12\gamma_1\gamma(x_0)^2\kappa A +15\gamma_1^2 A^2}
\end{equation}
and can be made arbitrarily small by formally 
varying the parameters $A$ and $\kappa$, 
meaning that one can in principle have two wide regimes of initial diffusive and subsequent ballistic dynamics. 
Two regimes with a crossover time $t_{1\rightarrow 2}$  already exist for equilibrium noise $\gamma_1=0$
 but the effect is persistent and tunable via  nonequilibrium noise as documented by Eq.(\ref{t_crossover}).

\begin{figure}
	\includegraphics[scale=0.21]{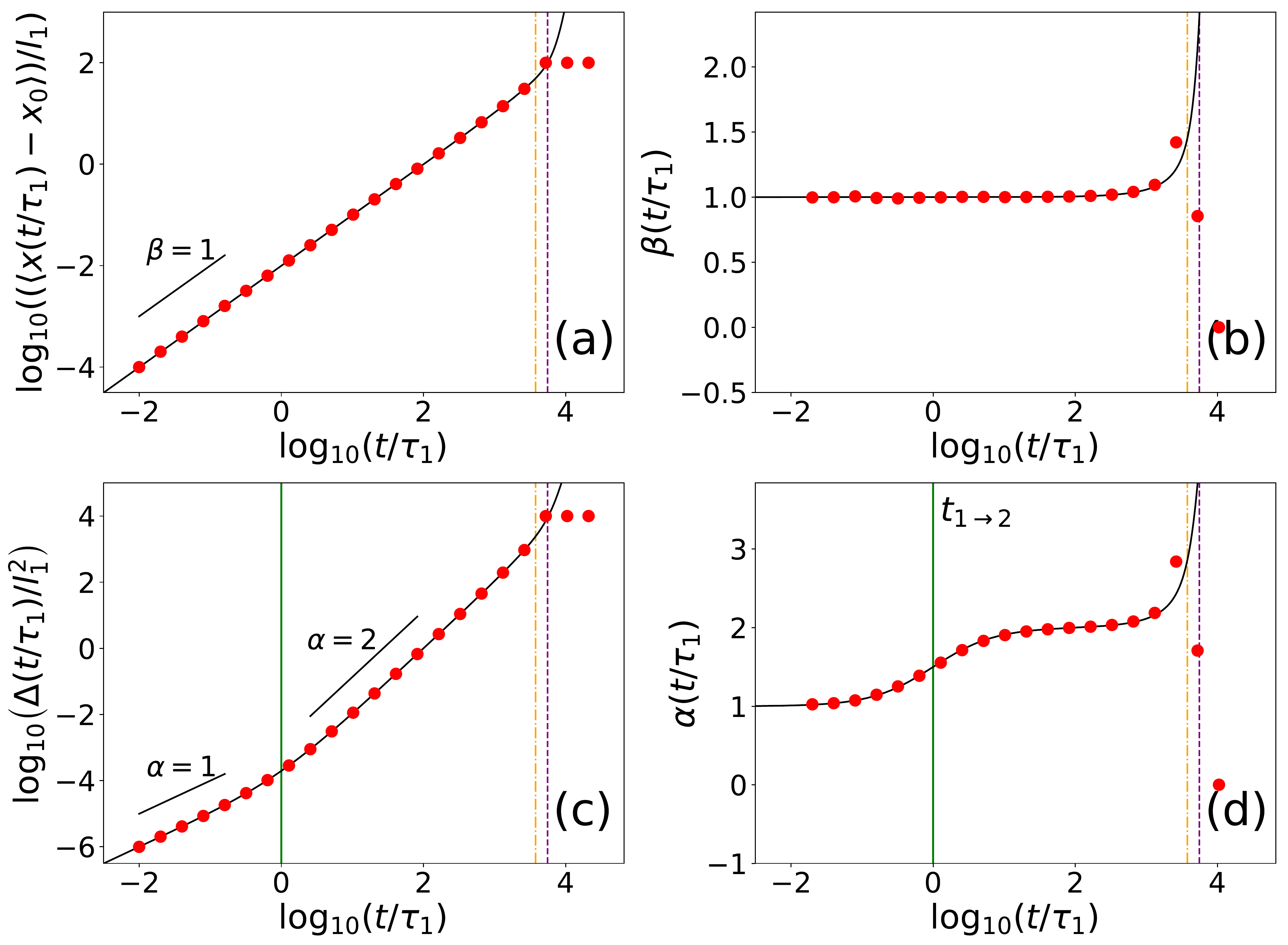}
\caption{Same as Fig. \ref{F2}, but now for $n=1$. (a) mean displacement; (b) scaling exponent $\beta(t)$; 
(c) mean-squared displacement $\Delta(t)$; (d) scaling exponent $\alpha(t)$. In (c) and (d) 
the crossing time $t_{1\rightarrow 2}$ is indicated by a vertical green line. 
Parameter values are: $ \kappa = \gamma_0l_1/\tau_1$, 
$x_0=100l_1$.  
}
\label{F3}
\end{figure}

Results for the MD and the MSD as well as the scaling exponents and passage times $t_1^c$ and $t_2^c$ are shown in Fig. \ref{F3},
obtained by both theory and simulation. 
The crossover between the initial diffusive and subsequent ballistic behavior in the MSD
is clearly visible, in particular in $\alpha(t)$, which shows a plateau around $\alpha(t)=2$ for intermediate times. The simulation data even reveal a transient
 subsequent superballistic behavior, which then falls off once the particle arrives at the origin, where
it decelerates due to the opposed friction gradient. Again, for times smaller than the passage duration, 
theory and simulation are in very good agreement. Finally, the reason why the agreement of theory and numerics in Fig. \ref{F3} b) is much better than that of Fig. \ref{F2} b) is that the drift is now dominated by the deterministic potential, while in the case of the free particle it was completely noise-driven.


\begin{figure}
	\includegraphics[scale=0.21]{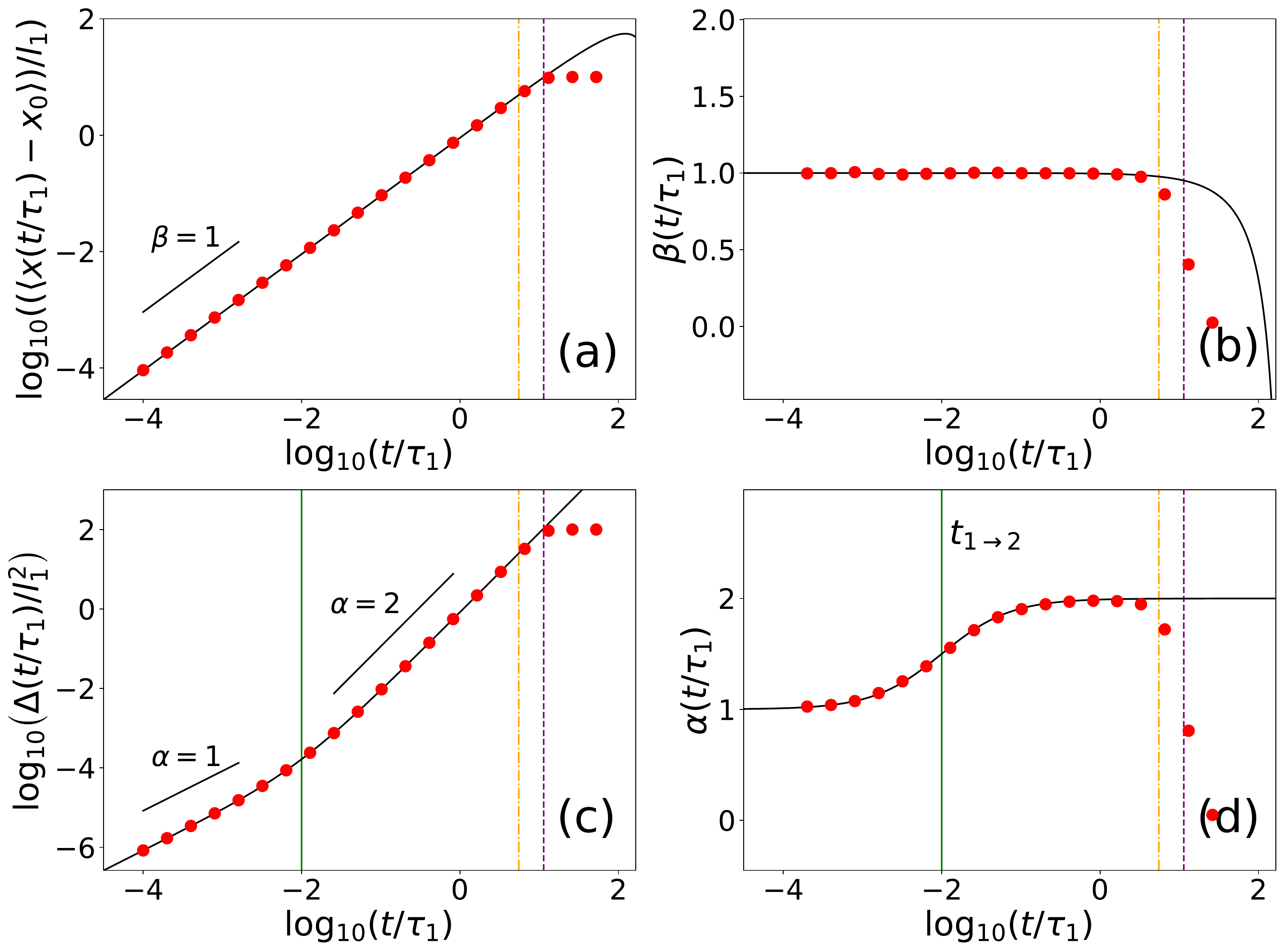}
	\caption{Same as Fig. \ref{F2}, but now for $n=2$: (a) mean displacement $\langle x(t)-x_0 \rangle$ and (b) scaling exponent $\beta(t)$; (c) mean-squared displacement $\Delta(t)$  and (d) scaling exponent $\beta(t)$ 
The parameters are  $ \kappa = \gamma_0/\tau_1$ and $x_0=10l_1$. }
         \label{F4}
\end{figure}

{\bf Harmonic potential.} Finally, for the harmonic oscillator: $U(x)=\frac{1}{2}\kappa x^2$, or
$n=2$, separation of variables is no longer possible and we therefore resort to a short-time expansion gained by
perturbation theory (see \cite{breoni_2020}). In doing so, first we take the solution of the $(p,n)=(1,1)$ system, with a constant force of $-\kappa x_0$, and next we consider a harmonic oscillator potential centered in $x_0$ as a perturbation. % the Brownian oscillator with equilibrium noise
%\begin{equation}
%(\gamma_0+\gamma_1x_0)\dot{x}_{BO}(t)=-\kappa x_{BO}(t) + \sqrt{A}\xi(t)
%\end{equation}
%and later we consider the constant friction gradient as a perturbation, such that $x(t)=x_{BO}(t)+h(t)$
%\begin{equation}
%\gamma_1(x_{BO}(t)-x_0)\dot{x}_{BO}(t)+(\gamma_0+\gamma_1 x_{BO}%(t))\dot{h}(t)\simeq 0.
%\end{equation}
 Following this procedure, the short-time expansions of the MD and MSD are:
\begin{align}
\langle x(t)-x_0 \rangle&=-\text{sgn}(x_0)\left(\left[\frac{\kappa |x_0|}{\gamma(x_0)}+\frac{\gamma_1 A}{2\gamma(x_0)^3}\right]t\right.\nonumber\\
&\left.+\left[ -\frac{|x_0| \kappa^2}{2\gamma(x_0)^2}+\frac{\gamma_1 \kappa^2 x_0^2 }{2\gamma(x_0)^3}	-\frac{3}{4}\frac{\kappa A\gamma_1}{\gamma(x_0)^4}\right.\right.\\
&\left.\left.+\frac{3}{2}\frac{|x_0|\kappa\gamma_1^2A}{\gamma(x_0)^5}+\frac{15}{8}\frac{\gamma_1^3A^2}{\gamma(x_0)^7}\right] t^2\right)+\mathcal{O}(t^3) \nonumber
\end{align}
and 
\begin{align}
\Delta(t)&=\frac{A}{\gamma(x_0)^2}t+\left[ \frac{x_0^2 \kappa^2}{\gamma(x_0)^2}	-\frac{\kappa A}{\gamma(x_0)^3}\right.\\
&+\left.3\frac{\gamma_1 \kappa |x_0| A}{\gamma(x_0)^4}+\frac{15}{4}\frac{\gamma_1^2A^2}{\gamma(x_0)^6}\right] t^2+\mathcal{O}(t^3)\, .\nonumber 
\end{align}
In this case, the MD only shows a linear behavior, while the MSD displays two different regimes, diffusive and ballistic, separated by the crossing time
\begin{eqnarray}
t_{1\rightarrow 2} & = & \\
&& \hspace*{-1.2cm}  \frac{4\gamma(x_0)^4A}{4\gamma(x_0)^4x_0^2\kappa^2-4\gamma(x_0)^3\kappa A+12\gamma_1\gamma(x_0)^2\kappa x_0 A+15\gamma_1^2A^2}. \nonumber
\end{eqnarray}
Fig. \ref{F4} shows the comparison of the  perturbation theory with the full numerical simulations revealing very good agreement for times smaller than a typical passage time. Clearly, for larger times, the particles becomes confined by the harmonic potential around the origin as signaled by a plateau arising in the MD and MSD for times larger than the typical passage time.
Correspondingly, both scaling exponents $\beta (t)$ and $\alpha (t)$  drop  to zero. \\

\subsection{Linear friction gradient}

\begin{figure}
	\includegraphics[scale=0.21]{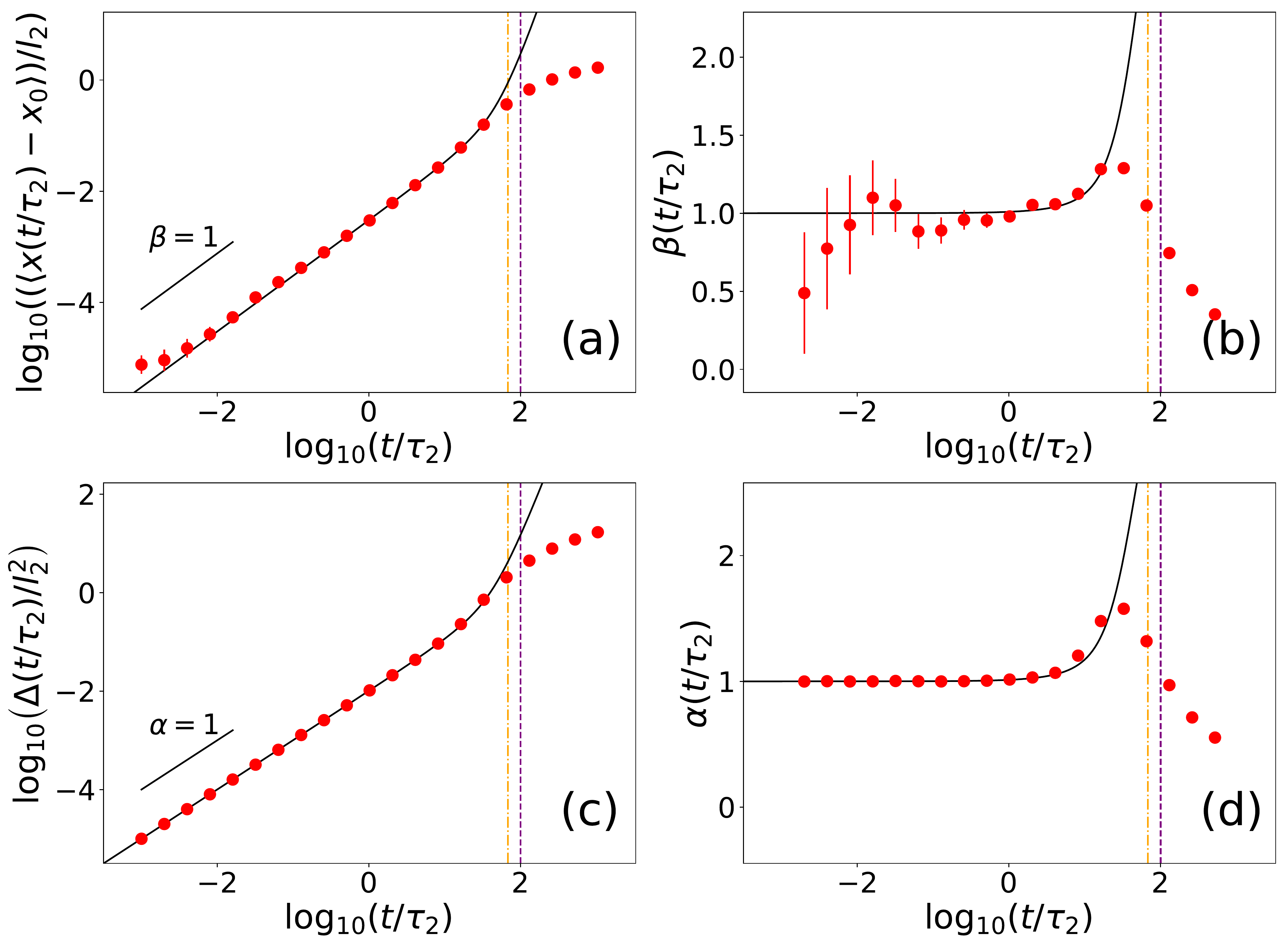}
	\caption{Linear friction gradient $p=2$ for a free particle ($n=0$): (a) mean displacement $\langle x(t)-x_0 \rangle$ and (b) scaling exponent $\beta(t)$; 
	(c) mean-squared displacement $\Delta(t)$, (d) scaling exponent $\alpha(t)$. The length units used is $l_2=\sqrt{\gamma_0/\gamma_1}$ and the time unit is $\tau_2=l_2^2/A$. The chosen initial position is $x_0=3l_2$.
 }
\label{F5}
\end{figure}

\begin{figure}
	\includegraphics[scale=0.21]{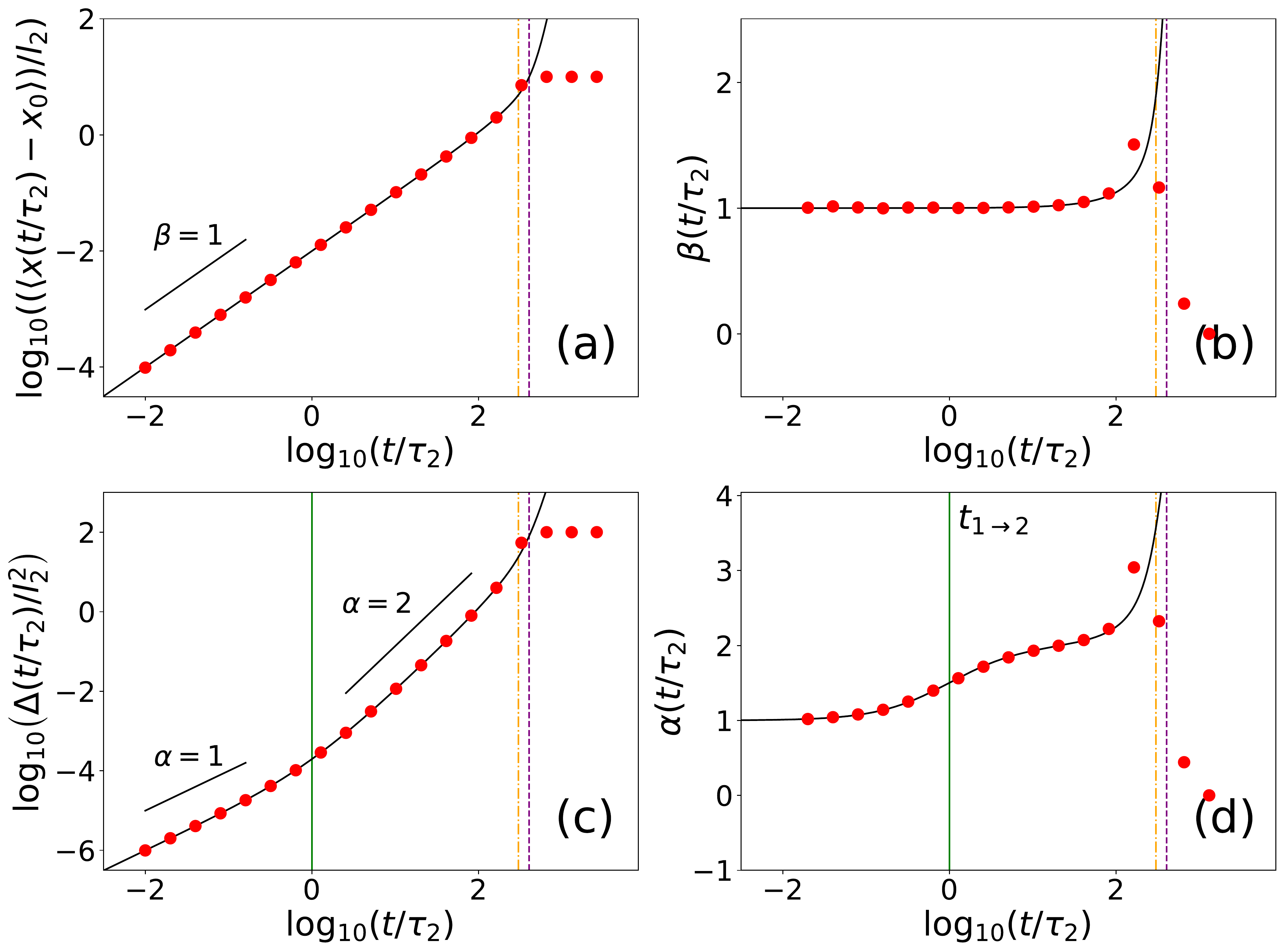}
	\caption{Same as Fig. \ref{F5}, but now for $n=1$: (a) mean displacement $\langle x(t)-x_0 \rangle$ and (b) scaling exponent $\beta(t)$; 
	(c) mean-squared displacement $\Delta(t)$, (d) scaling exponent $\alpha(t)$ and indicated crossing time $t_{1\rightarrow 2}$.
Parameter values: $\kappa=\gamma_0l_2/\tau_2$, $x_0=10l_2$.
}
\label{F6}
\end{figure}

We now turn to a linear friction gradient, $p=2$, where there is no nonanalyticity 
in the spatial dependence of the friction at the origin.
Then Eq.(\ref{active}) becomes
\begin{equation}
(\gamma_0 + \gamma_1 x^2) \dot{x}(t) = -U'(x) + \sqrt{A} \xi(t). 
\end{equation}
%We restrict ourselves in this section to the free case and the one with potential $U(x)=\kappa|x|$. 
%comment HL: Davide, why is perturbation theory not possible, can we get anything for n=2?
Bearing in mind that the free case is a simple special case of the $n=1$ one (for $\kappa=0$), we directly show the results for $n=0,1$ for any $\kappa\geq0$. The MD is
\begin{eqnarray}
\label{MD21}
\langle x(t) -x_0 \rangle & = & \sum_{m=1}^{\infty}a_m\langle\zeta^m(t)\rangle \\
&& \hspace*{-2cm} =  \sum_{m=1}^{\infty}a_m\sum_{k=0}^{\lfloor m/2 \rfloor}\frac{m!A^k(-\text{sgn}(x_0)\kappa)^{m-2k}}{(m-2k)!2^kk!}t^{m-k}\, , \nonumber
\end{eqnarray}
where the factors $a_m$ are straightforwardly 
obtained by Taylor expanding the expression $(x(t)-x_0)$, calculated using separation of variables, in powers of 
\begin{equation}
\zeta(t) = - \text{sgn}(x_0)\kappa t + \sqrt{A}\int_0^t dt' \xi(t') \, .
\end{equation} 
Here $a_1=\gamma(x_0)^{-1}$, but the expressions for the coefficients $a_m$ for $m\geq 2$ are quite involved so that we refrain from showing them explicitly.
In a similar way, the MSD is
\begin{eqnarray}
\label{MSD21}
\Delta(t) & = & \sum_{m=2}^{\infty}b_m\langle\zeta^m(t)\rangle\\
& = & \sum_{m=2}^{\infty}b_m\sum_{k=0}^{\lfloor m/2 \rfloor}\frac{m!A^k(-\text{sgn}(x_0)\kappa)^{m-2k}}{(m-2k)!2^kk!}t^{m-k},\,  \nonumber
\end{eqnarray}
where $b_2=\gamma(x_0)^{-2}$ and the coefficients $b_m$ for $m \geq 3$ are again quite involved. The behavior of both the MD and the MSD are very similar to the ones for the $p=1$ case, with a simple diffusive behavior if $\kappa=0$ and 
both a diffusive and ballistic behavior otherwise.
A comparison between theory and simulations is shown in Fig. \ref{F5} for the free case and in Fig. \ref{F6} for $n=1$.

\begin{figure}
	\includegraphics[scale=0.21]{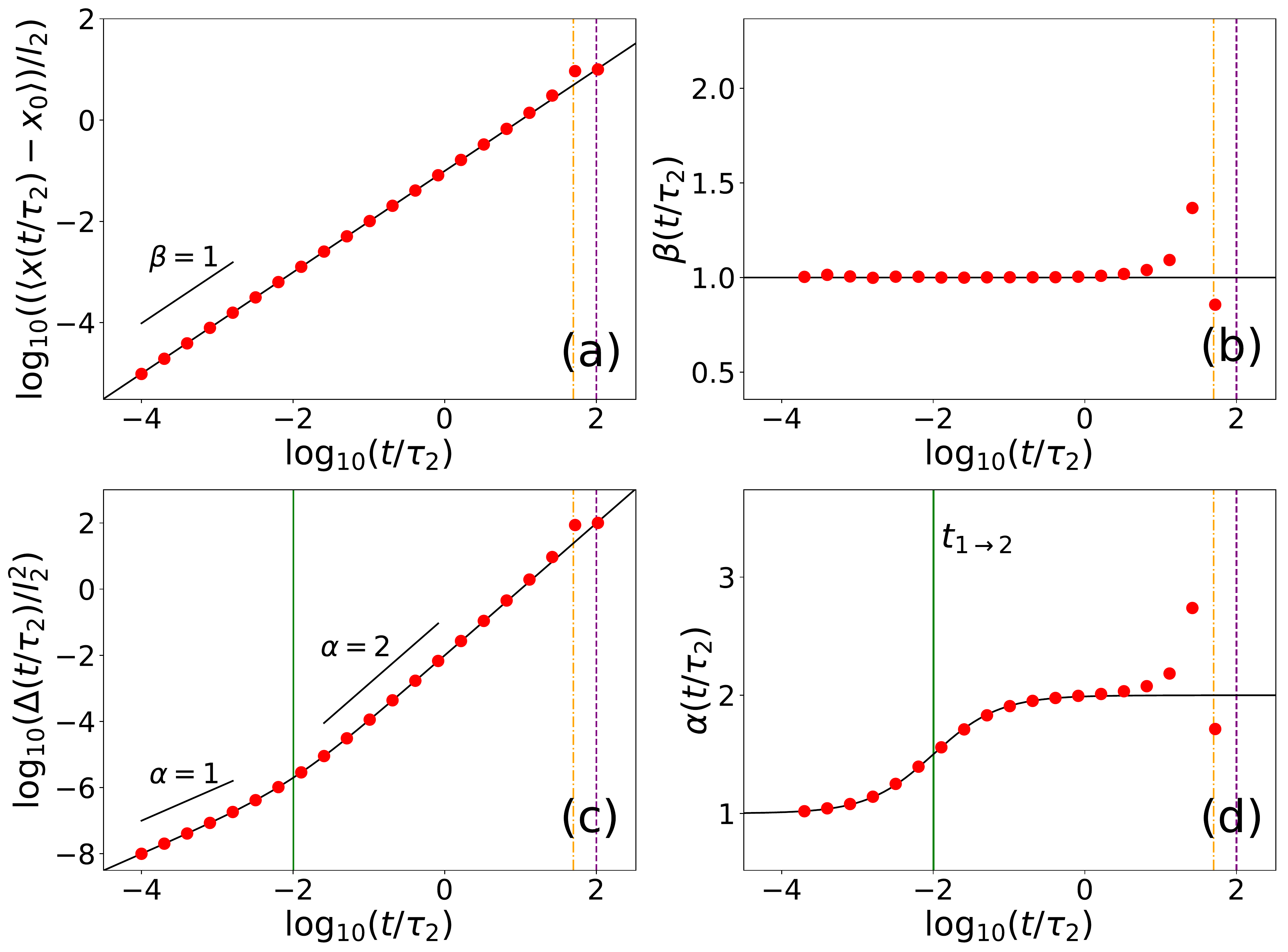}
	\caption{Same as Fig. \ref{F5}, but now for $n=2$: (a) mean displacement $\langle x(t)-x_0 \rangle$ and (b) scaling exponent $\beta(t)$; 
	(c) mean-squared displacement $\Delta(t)$, (d) scaling exponent $\alpha(t)$ and indicated crossing time $t_{1\rightarrow 2}$.
Parameter values: $\kappa=\gamma_0/\tau_2$, $x_0=10l_2$.
}
\label{F7}
\end{figure}

For the case $n=2$ we used perturbation theory to calculate up to the first order in time for the MD and up to the second order in time for the MSD:
\begin{equation}
\langle x(t)-x_0 \rangle=\left(-\frac{\kappa x_0}{\gamma(x_0)}+a_2 A\right) t+\mathcal{O}(t^2),
\end{equation}
\begin{eqnarray}
\Delta(t)&=& \frac{ At}{\gamma(x_0)^2} + \\ 
&&\hspace*{-1cm}\left[\frac{\kappa^2 x_0^2}{\gamma(x_0)^2}-\frac{\kappa A}{\gamma(x_0)^3}-3b_3 A \kappa x_0 + 3b_4 A^2\right]t^2+\mathcal{O}(t^3),\nonumber
\end{eqnarray}
where the $a_i$ and $b_i$ are the coefficients already used in Eqs. (\ref{MD21}) and (\ref{MSD21}). We see again a linear behavior for the MD while the MSD goes from diffusive to ballistic. In Fig. \ref{F7} we compare these results with numerical simulations.

\section{Long-time behavior} 

We now consider the stationary long-time behavior. In order to keep a normalized probability distribution function, we
confine the system in a potential $(n=1,2)$. The stochastic process then admits a stationary PDF on the infinite line in the $x$-coordinate 
which can be computed from the Fokker-Planck equation corresponding to the process Eq.(\ref{active}).
We rewrite, analogous to Eq.(\ref{active2}),
\begin{equation} \label{langevin2} 
\dot{x}(t) = a(x) + b(x) \xi(t)
\end{equation}
with
\begin{equation}
a(x) \equiv - \frac{U'(x)}{\gamma(x)}\,,\,\,\, b(x) \equiv \frac{\sqrt{A}}{\gamma(x)}\, .
\end{equation}
The Fokker-Planck equation for this case has been derived in \cite{ryter_brownian_1981,baule_exact_2008} and reads as
\begin{equation} \label{fp}
\partial_t p(x,t) = -\partial_x[a(x)p(x,t)] + \frac{1}{2}\partial_x[b(x)[\partial_x[b(x)p(x,t)]]]\, ,
\end{equation}
\\
admitting a stationary solution at zero flux which is given by
\begin{equation}  \label{pdf}  
p(x) = \frac{N}{b(x)} \exp \left[\int^x dy\, \frac{2a(y)}{b^2(y)}\right],
\end{equation}
where $N$ is a normalization factor. The integrand in the exponential of Eq.(\ref{pdf}), denoted by $I(y)$, can be expressed in terms of the confining potential and
the friction term as
\begin{equation} \label{Iy}
I(y)= - \frac{2}{A} U'(y) \gamma(y) 
\end{equation}
which shows that it is given by polynomial expressions for the cases we address now.

\begin{figure}
	\includegraphics[scale=0.335]{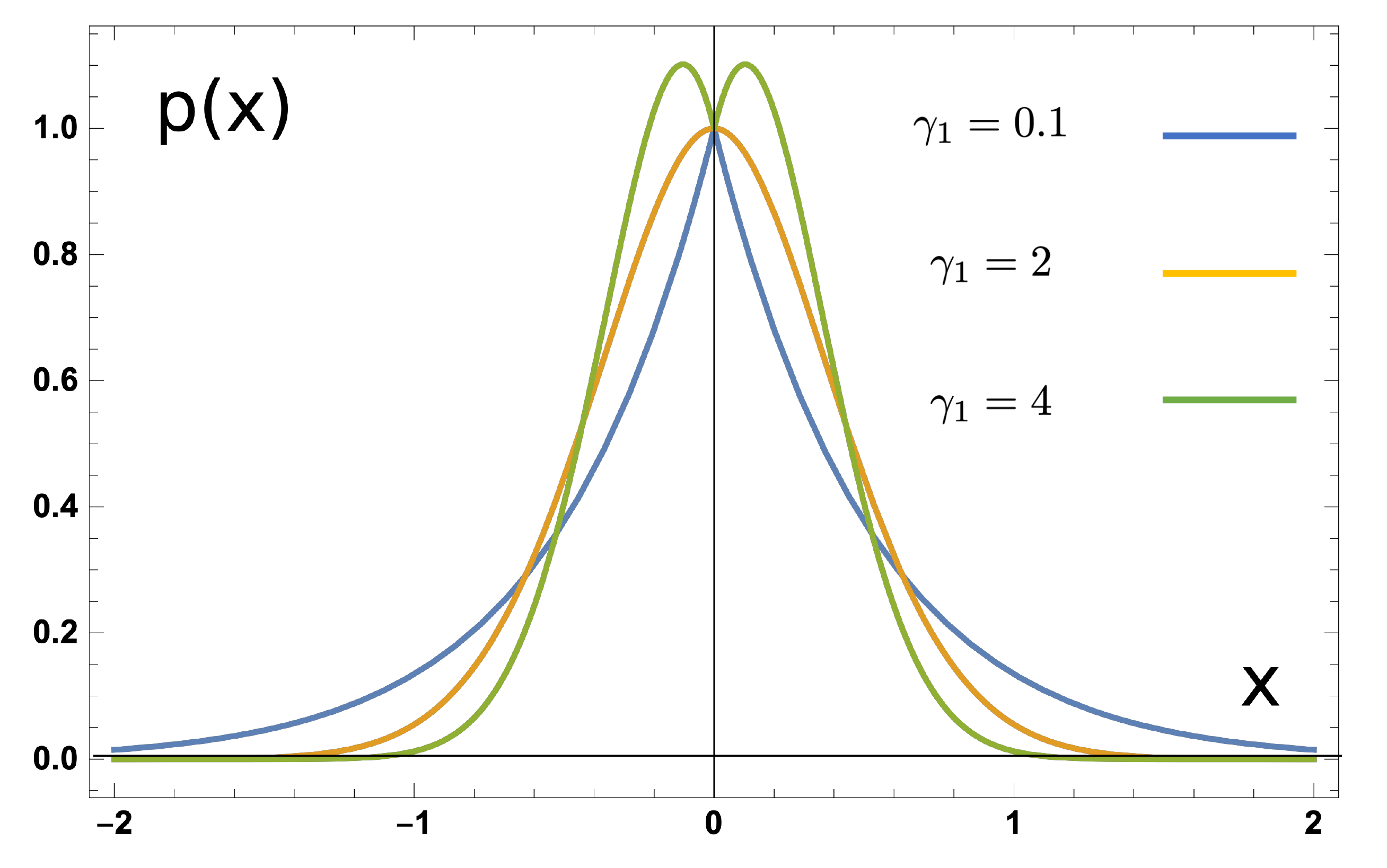}
	\caption{Normalized PDF $p(x)$ for $\gamma(x) = \gamma_0 + \gamma_1 |x| $ and $U(x) = \kappa |x|$, hence $(p,n) = (1,1)$. Shown are curves 
	for three sets of values of $\gamma_0 = 1$ with all other parameters set to numerical values of one. 
  $\gamma_1 = 0.1$ blue curve, Laplace-distribution; 
	$\gamma_1 = 3$; yellow curve, Gaussian distribution. With $\gamma_1 = 4 $ one obtains a 
	bimodal ``mirrored" Gaussian curve.}
         \label{F8}
\end{figure}

Taking $ \gamma(x) = \gamma_0 + \gamma_1 |x|^p$ and $U(x) = (\kappa/n) |x|^n $, which covers both our cases of interest for
$ p = 1, 2 $, $ n = 1,2 $, one obtains from Eq.(\ref{pdf}) 
\begin{eqnarray}  \label{pdf-2}  
p(x) & = & \frac{N}{\sqrt{A}} (\gamma_0 + \gamma_1 |x|^p) \times \\
&& \exp \left[-\frac{2\kappa}{A} 
\left( \frac{\gamma_0}{n} |x|^n  + \frac{\gamma_1}{n+p} |x|^{n+p} \right)\right]\, . \nonumber
\end{eqnarray}
We can now discuss the different cases as a function of the exponent pairs $(p,n)$. For the lowest-order case
$ (p,n) = (1,1)$ one has the superposition of the exponentials of a Laplace- and a Gaussian distribution, as shown in Fig. \ref{F8}.
\begin{figure}
	\includegraphics[scale=0.33]{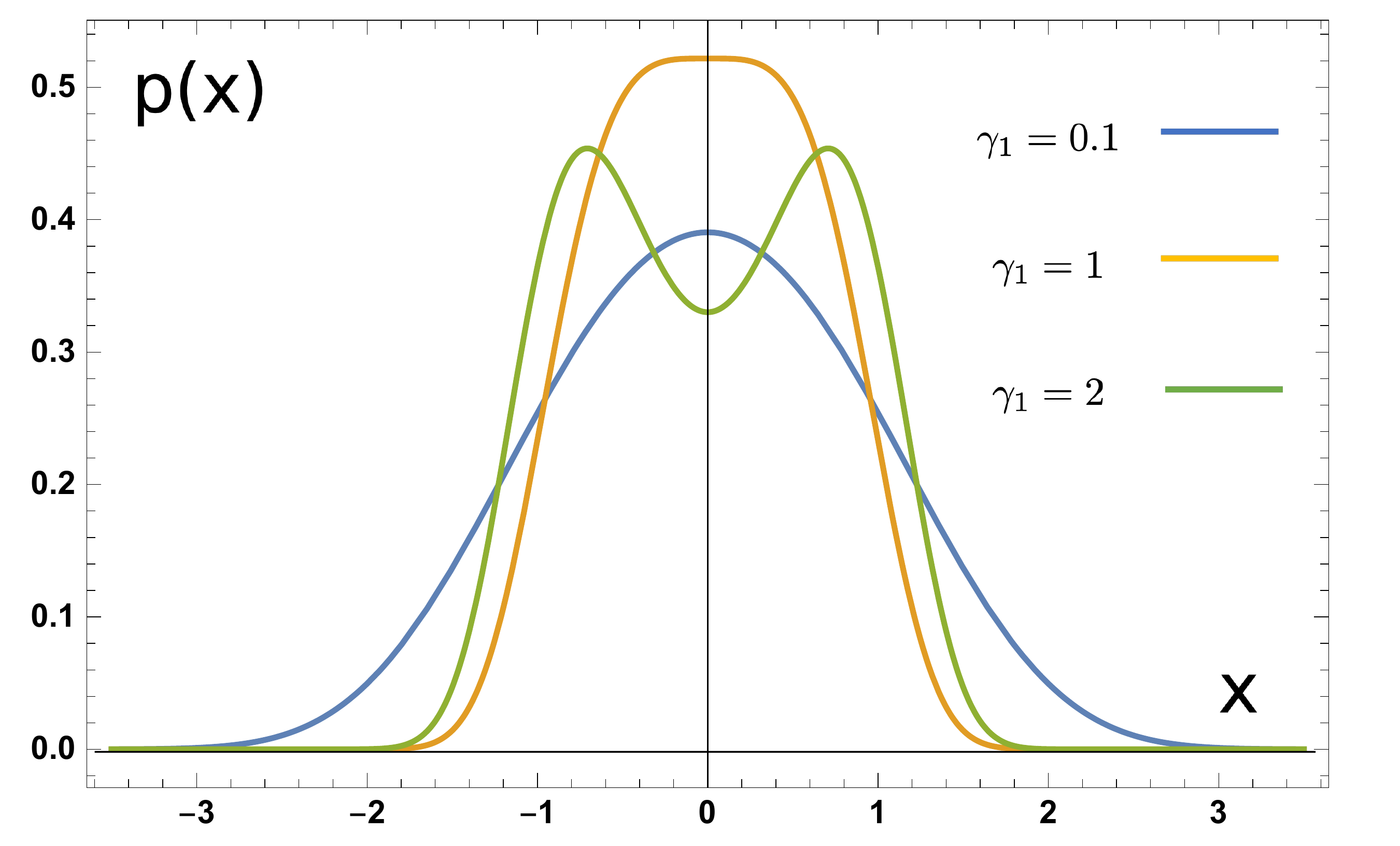}
	\caption{Case $(p,n) = (2,2)$. Normalized PDF $p(x)$  for $U(x) = \kappa x^2/2$ for three sets of values of $\gamma_1 $ with all other parameters set to 1.
	$\gamma_1 = 0.1$:  Gauss-like-distribution; $\gamma_1 = 1$: flat-top distribution; $\gamma_1 = 2$; bimodal Gaussian-like distribution.}
        \label{F9}
\end{figure}	
The resulting PDF therefore interpolates between a Laplace-like distribution in the limit $ \gamma_0 \gg \gamma_1$ and a Gaussian-like distribution 
up to $\gamma_1 = 2\gamma_0^2 a $, where the coefficient $a \equiv \kappa/A$ takes care of the different physical dimensions of $\gamma_0$ and $\gamma_1$;
we set $a \equiv 1$. For still larger values of $\gamma_1 \gg \gamma_0 $, the monomodal Gaussian distribution splits in what 
we call a bimodal ``mirrored" Gaussian distribution. This name reflects the observation that the resulting distribution looks like a Gaussian placed close to a mirror, with 
the parts of the image behind the mirror cut out. It is important to note that for the presence of these different distribution forms the friction-dependent 
prefactor is important; at $x = 0$ it is a constant, but within a range of $x$-values around zero it reweights the distribution away from that constant, 
before for large values of $x$ the exponential contribution becomes dominant. 

The PDF in the case $(p,n) = (2,1) $ shows the same behavior, which can be read off from the exponents. The leading Laplacian terms in
unaltered since $ n = 1 $, while the subsequent term now acquires a cubic nonlinearity.
In the case $ (p,n) = (1,2) $ the leading order term is now a Gaussian term, which therefore dominates at small values of $\gamma_1$. As in the previous cases,
for increasing values of $\gamma_1$, the distribution immediately turns into a mirrored Gaussian-distribution, i.e. the maximum of the distribution splits into two
maxima. 

Finally, $(p,n) = (2,2)$ the polynomial in the exponent is even and of fourth-order, with Gaussian behavior dominating at low values 
of $\gamma_1$. Going from small to large $\gamma_1$, one now crosses over from a Gaussian-like to a bimodal Gaussian-like-distribution, 
which now is smooth at $x=0$ due to the absence of modulus terms. This form is shown in Fig. \ref{F9}. All behaviors found are summarized in
Table I. 
\begin{table}[htp] 
\caption{Graphic summary of the PDFs for the cases $(p,n)$ for $ p = 1, 2$, $ n = 1, 2$, varying only the friction strengths $\gamma_0$ and
$\gamma_1$. For $\gamma_1 = 0$, the
distributions are either Laplacian (L) or Gaussian (G); left-most points. Increasing $\gamma_1$ leads to mirrored Gaussian behavior (MG),
passing via Gaussian behavior at $\gamma_1 = 2\gamma_0^2 a $, $a = (\kappa/A) \equiv 1$. This applies to both $(p,n) = (1,1)$ and $(p,n) = (2,2)$. For $(p,n) = (1,2)$,
starting from a Gaussian at $\gamma_1 = 0$, the PDF changes into a mirrored Gaussian shape for finite positive $\gamma_1$. Finally,
for $(p,n) = (2,2)$, Gaussian behavior changes into bimodal Gaussian behavior (BG), passing via a flat-top behavior (FT) at $\gamma_1 
= \gamma_0^2 a $, with again $a = 1$.} 
\begin{center}
\begin{tabular}{|c|c|}
\hline 
$(p,n)$ & \\
& $\sbullet[1.5]$ \hspace{-1.5mm} $\xrightarrow{\makebox[4cm]{0 \hspace{3.6cm} $\gamma_1$}}$\\
\hline 
$(1,1)$ & \\
& $\sbullet[1.5]$ \hspace{-1mm}$\xrightarrow{\makebox[1.9cm]{L \hspace{1.3cm} G}}$\hspace{-2.5mm} {$\sbullet [1.5]$}\hspace{-2mm} $\xrightarrow{\makebox[2.1cm]{\hspace{13mm} MG}}$\\
$(2,1)$ & $\gamma_1 = 2\gamma_0^2$ \\
\hline  
& \\
$(1,2)$ 
& $\sbullet[1.5]$ \hspace{-1mm}$\xrightarrow{\makebox[4.2cm]{G \hspace{3.2cm} MG}}$ \\
& \\
\hline
& \\
$(2,2)$ & $\sbullet[1.5]$ \hspace{-1mm}$\xrightarrow{\makebox[0.9cm]{G \hspace{0.2cm} FT}}$\hspace{-2.5mm} {$\sbullet[1.5]$} \hspace{-2mm} 
$\xrightarrow{\makebox[3cm]{\hspace{24mm} BG}}$ \\
& \hspace{-2.2cm} $\gamma_1 = \gamma_0^2$ \\
\hline
\end{tabular}
\end{center}
\end{table}

We end by considering the robustness of our results with respect to thermal fluctuations. Following Baule {\it et al.} \cite{baule_exact_2008}, we consider the
Langevin equation
\begin{equation}
\gamma(x)\dot{x}(t) = -U'(x) + \sqrt{2\gamma(x) k_BT} \eta(t) + \sqrt{A} \xi(t)  
\end{equation}
where $\eta(t)$ is a Gaussian white noise. The thermal and active processes $\eta(t)$ and $\xi(t)$ being uncorrelated, they
can be superimposed to $ \xi_T(t) = \eta(t) + \xi(t) $, leading to
\begin{equation}
\dot{x}(t) = a_T(x) + b_T(x)\xi_T(t)\, ,
\end{equation}
with 
\begin{equation}
a_T(x) = - \frac{1}{\gamma(x)}\left(U'(x) + \frac{k_BT}{2}\frac{\gamma'(x)}{\gamma(x)}\right)
\end{equation}
and
\begin{equation}
b^2_T(x) = \frac{1}{\gamma(x)} \left(2k_BT + \frac{A}{\gamma(x)}\right)\, .
\end{equation}

The integrand $I(y)$ in the exponential of the PDF reads as
\begin{equation}
I(y) = - 2\left[\frac{U'(y)\gamma(y) + k_BT \gamma'(x)}{A + 2k_BT\gamma(x)}\right]\, ,
\end{equation}
which can be compared with Eq.(\ref{Iy}) in the purely active case. As a robustness check 
it suffices to examine the behavior of the integrand $I(y)$ near the origin for small values of $y$ and for $y \rightarrow \pm \infty$,
for our four cases $(p,n)$, $ n = 1,2 $, $p = 1,2 $. For the behavior near the origin one finds that the dominator behaves in
a similar fashion as $I(y)$ of Eq.(\ref{Iy}), generating a polynomial with identical powers, since the temperature-dependent
term either contributes a $\mbox{sgn}(x)$ for $p =1$ or a linear term for $p = 2$. The qualitative behavior of the PDFs remains
thus unaltered. For large arguments, one sees that generally $I(y)$ behaves as
\begin{equation}
I(y) \propto -\frac{U'(y)}{k_BT}\, ,
\end{equation}
such that the tails of the distributions are determined by thermal fluctuations and decay exponentially, i.e. Laplace-like for $ n = 1 $ or
Gaussian-like for $ n = 2 $; the active noise and the friction term then only play a role in the prefactor of the PDF.

\section{Discussion and Conclusions} 

In this work we have studied the stochastic dynamics of an active-noise driven
 particle  under the influence of 
a space-dependent friction  and confinement. In order to elucidate the effect of the space-dependence of the friction term,
we start the dynamics for large initial values, so that the friction term dominates the dynamics. For the case of a free particle, 
a particle running down a ramp and a harmonic potential we have determined the mean displacement and mean-squared displacement 
and the corresponding scaling exponents $\beta(t) $ and $\alpha(t)$ in a short-time expansion. The mean displacements generally show 
diffusive behaviors, while a crossover to a ballistic regime is observed for the mean-squared displacement, except for the free particle case.

Further, we have determined the effect of the friction term in the presence of a confining potential $U(x) \propto |x|^n $ for $n = 1,2$
for long times. We have analytically computed the stationary probability density functions from the Fokker-Planck equation.  
These solutions can be classified according to the exponent pairs $(p,n)$ and the relative magnitude of the friction 
coefficients $\gamma_0 $ and $\gamma_1$. One observes that the friction law and the confinement
potential conspire to generate a set of generic behaviors: Laplace-like and Gaussian-like distributions for $ n = 1$ and $n = 2$, respectively,
if the spatially-dependent friction term is small ($\gamma_1 \ll \gamma_0 $); this behavior crosses over for  $\gamma_1 = \gamma_0^2$ 
to Gaussian behavior for both $ p = 1, 2$. In the case of $ n = 2$, Laplace-like behavior is absent. For all cases of $(p,n)$ with $ n=1,2 $
$ p = 1,2$, one observes that for $\gamma_1 \gg \gamma_0$, the stationary PDF displays a mirrored or bimodal Gaussian-like behavior. 
Therefore, generally for all combinations of $(p,n)$, at sufficiently strong space-dependent friction, the PDF becomes a bimodal distribution
with a symmetrically increased weight off-center of the potential minimum. 

To conclude, our study extends current studies on active particles in one dimension by the inclusion of a space-dependent friction and therefore
links the problem to earlier studies of molecular motors on linear tracks. Investigations of the stationary probability density functions for the
run-and-tumble process have already generated an extended catalog of distributions, see, e.g. \cite{dhar_run-and-tumble_2019}, in 
which also bimodal-type PDFs appear (see their Fig.7), or \cite{sevilla19}. Placed in this context, the present study reveals a basic classification method in 
which such complex distributions are categorized for the case of a space-dependent friction. Our model system allows to extract the 
mechanism of shape change of the PDFs in a particularly clear manner. 

Our theory can be extended to active noise driven motion in two spatial dimensions. A special two-dimensional example is given by a radially symmetric situation, where the friction $\gamma$ solely depends on the radial distance $r=\sqrt{x^2+y^2}$. This case can be solved with similar methods as proposed in this paper. Another possible extension of our model could treat full viscosity landscapes \cite{viscosity_landscape1,viscosity_landscape2,viscosity_landscape3,viscosity_landscape4}. Moreover inertial effects can be included
in the particle dynamics \cite{Ldov,Loewen_JCP,Sprenger,Caprini}. Finally collective effects for many active-noise driven particles such as motility-induced phase separation should be explored \cite{MIPS1,MIPS2}.
\\

{\bf Acknowledgement.} DB is supported by the EU MSCA-ITN ActiveMatter, (Proposal No. 812780). RB is grateful to HL for the invitation to a stay at the Heinrich-Heine University in 
D\"usseldorf where this work was performed.

\begin{center}
\bf{Appendix: Analytical and Numerical calculations}
\end{center}

\flushleft{\bf Calculation of the mean displacement in the $(p,n)=(1,0)$ case}
\\

In this appendix we show how we obtained Eq.(\ref{Eq10}) starting from Eq.(\ref{quadratic}). First, we notice that Eq.(\ref{quadratic}) can be written as:
\begin{equation}
x(t) = - \frac{\gamma_0}{\gamma_1} + \frac{\gamma_0 + x_0\gamma_1}{\gamma_1}\sqrt{1+ \frac{2\gamma_1\sqrt{A}}{(\gamma_0 + x_0\gamma_1)^2}\int_0^tdt'\xi(t')}.  
\end{equation}
Given $x(t)$, we can write the equation for the MD:
\begin{align}
&\langle x(t) - x_0 \rangle  \\ 
&=\frac{\gamma_0 + x_0\gamma_1}{\gamma_1}\left(\left\langle\sqrt{1+ \frac{2\gamma_1\sqrt{A}}{(\gamma_0 + x_0\gamma_1)^2}\int_0^tdt'\xi(t')}\right\rangle-1\right).  
\end{align}
To calculate the average in this expression, we have to Taylor expand the square root, using the following formula:
\begin{equation}
\sqrt{1+a}=1+\frac{a}{2}+\sum_{m=2}^{\infty} (-1)^{m-1}\frac{(2m-3)!}{2^{2m-2}m!(m-2)!}a^{m},
\end{equation}
where we substitute
\begin{equation}
a\equiv \frac{2\gamma_1\sqrt{A}}{(\gamma_0 + x_0\gamma_1)^2}\left\langle\int_0^tdt'\xi(t')\right\rangle.
\end{equation}
Eq.(\ref{Eq10}) follows directly.

{\flushleft{\bf Numerical treatment of the Langevin equation}}
\\

The stochastic equation
\begin{equation} \label{langevin2} 
\dot{x}(t) = a(x) + b(x) \xi(t)
\end{equation}
is of the standard form
\begin{equation}
\label{stoceq}
dx_t=a(x_t)dt+b(x_t)dW_t,
\end{equation}
where $W_t$ represents a Wiener process. In order to solve this equation numerically in the Stratonovich paradigm, we implement a predictor-corrector scheme. 
In such a scheme, one first performs a full time step evolution of the position of the particle  $x(t_i)$ using the same time coefficients $a(x(t_i))$ and $b(x(t_i))$. 
This predicted position $x_p$ is used to calculate $a(x_p)$ and $b(x_p)$ and proceed to finally calculate the position at time step $t_{i+1}$ using the averages of the coefficients calculated for $x(t_i)$ and $x_p$. To implement the Stratonovich paradigm, using this kind of average only for the stochastic part (and hence the $b(x)$) is necessary, but we preferred to apply this procedure as well to the deterministic part in order to improve stability of the result.
The method we decided to use for the time evolution is thus a Milstein scheme, of order $\mathcal{O}\left(\Delta t\right)$ \cite{milshtejn_approximate_1975}.  
The Milstein evolution of Eq.\ref{stoceq} can be written as:
\begin{eqnarray}
x(t_{i+1}) & = & x(t_{i})+a(x(t_{i}))\Delta t +  \\
&& \hspace*{-1.5cm} b(x(t_{i}))\Delta W(t_i)+\frac{1}{2}b(x(t_{i}))\frac{db(x(t_{i}))}{dx}((\Delta W(t_i))^2-\Delta t)\,, \nonumber
\end{eqnarray}
where $\Delta W(t_i)=W(t_{i+1}-W(t_i)$ is a normal-distributed random variable.

It should be noted that the fact that the Milstein scheme uses the derivative of the function $b(x)$, which for our model is discontinuous at $x = 0$ 
for the case $ p = 1$. This can be treated by adopting an algorithm developed in \cite{perez10}, employing colored noise from the Ornstein-Uhlenbeck process.

\end{document}